\documentclass[conference,a4paper]{IEEEtran}

\usepackage[cmex10]{amsmath}
\usepackage[cmex10]{amsmath}
\usepackage{array}
\usepackage{verbatim}
\usepackage{tikz}
\usepackage{epstopdf}
\usepackage{amsfonts}
\usepackage{amsmath}
\usepackage{amssymb}
\newtheorem{defi}{Definition}
\newtheorem{theorem}{Theorem}

\newtheorem{remark}{Remark}
\newtheorem{corollary}{Corollary}

\newcommand{\modn}{\hspace{-0.1in}\mod\Lambda}

\newcommand{\modnml}{\hspace{-0.05in}\mod\Lambda^m_{\ell}}

\newcommand{\modmprimet}{\hspace{-0.05in}\mod\Lambda^J_{m^{\prime \prime}3}}
\newcommand{\modmprimes}{\hspace{-0.05in}\mod\Lambda^J_{m^{\prime}2}}

\usepackage{cite}
\usepackage[center]{caption}
\usepackage{graphicx}
\usepackage{multicol}
\usetikzlibrary{shapes.misc}
\tikzset{cross/.style={cross out, draw=black, minimum size=2*(#1-\pgflinewidth), inner sep=0pt, outer sep=0pt},
cross/.default={1pt}}

\begin{document}
\title{Preserving Confidentiality in The Gaussian Broadcast Channel Using Compute-and-Forward}
\author{\IEEEauthorblockN{Parisa Babaheidarian$^1$, Somayeh Salimi$^2$, and Panos Papadimitratos$^3$}
\IEEEauthorblockA{
$^1$Boston University, $^2$Uppsala University, $^3$KTH Royal Institute of Technology}}

\date{}
\maketitle
\begin{abstract}
We study the transmission of confidential messages across a wireless broadcast channel with $K>2$ receivers and $K$ helpers. The goal is to transmit all messages reliably to their intended receivers while keeping them confidential from the unintended receivers. We design a codebook based on nested lattice structure, cooperative jamming, lattice alignment, and i.i.d. coding. Moreover, we exploit the asymmetric compute-and-forward decoding strategy to handle finite SNR regimes. Unlike previous alignment schemes, our achievable rates are attainable at any finite SNR value. Also, we show that our scheme achieves the optimal sum secure degrees of freedom of $1$ for the $K$-receiver Gaussian broadcast channel with $K$ confidential messages and $K$ helpers.
\end{abstract}
\section{Introduction}
Physical-layer security has been widely studied under different communication scenarios. The encoding strategies have been used to analyze these security scenarios can be grouped in two main categories: i.i.d. random coding and structured coding. Several achievability schemes were proposed within the first category. Csiszar and Korner discussed transmitting a confidential message over a broadcast channel with one legitimate receiver and one passive eavesdropper \cite{csiszar1978broadcast}; capacity results were obtained for a less noisy channel using random i.i.d. codes. Recently, several works in the second category shed lights on the advantage of structured codes in achieving security. In the absence of capacity results for different communication channels in general cases (i.e., no assumption on degradedness or specific channel gains), researchers have studied the secure degrees of freedom (s.d.o.f.) in the infinite SNR regime. Despite the promising performance that Gaussian i.i.d. codes show in maintaining reliability in AWGN channels, studies show that they achieve \textit{zero} sum \textit{secure} degrees of freedom \cite{xie2013secure} and \cite{he2014providing}.  In contrast, structured codes attain a positive secure degrees of freedom \cite{bagherikaram2010secure}, \cite{xie2014secure}, and \cite{babaheidarian2}. In \cite{xie2014secure}, a collection of one-hop communication scenarios were considered including the wiretap Gaussian broadcast channel with helpers. Xie and Ulukus in \cite{xie2014secure} and \cite{xieJournal2} suggested an achievable scheme for the considered security scenarios which was based on real alignment encoding, cooperative jamming, and maximum likelihood decoder which operated in the infinite SNR regime; they showed that following their schemes, optimal sum secure degrees of freedom are achievable. Also, in \cite{he2009secure} and \cite{he2014providing}, a lattice-based scheme was proposed for the Gaussian wiretap channel with one helper which was optimal at infinite SNR for a subset of channel gains.\\
\indent A lattice-based framework known as the \textit{compute-and-forward} framework \cite{nazer2011compute} was proposed to handle interference which enabled the decoder to decode integer linear combinations of the transmitted codewords. In \cite{babaheidarian2} and \cite{babaheidarian2016security}, we investigated the Gaussian wiretap multiple-access channel and the two-user Gaussian interference channel with confidential messages, respectively. For these models, we introduced achievable schemes which, unlike previous works, could operate at any finite SNR value and for almost all (real) channel gains. Furthermore, we derived constant gap results for the sum secure capacity. In this paper, we study the Gaussian broadcast channel with $K$ receivers and $K$ helpers for $K>2$. The case of $K=2$ with one helper was studied in \cite{xie2014secure} and optimal s.d.o.f. was obtained. \\
\indent In our model, the transmitter has an independent message for each receiver which needs to be kept confidential from other receivers. A set of $K$ helpers implicitly cooperate with the transmitter by sending out jamming signals with proper beam-forming, to assist the transmitter in preserving the confidentiality of messages at the unintended receivers. We propose an achievability scheme which works at any SNR value and for almost all real-valued channel gains\footnote{Except for a set of channel gains with small Lebesgue measure.}. We offer a set of lower bounds on individual secure rates and show that the sum secure rate is within a constant gap from the sum secure capacity for this channel model. Our achievable scheme combines the idea of jamming signals and beam-forming with asymptotic alignment in \cite{xieJournal2} and a generalization of the compute-and-forward framework in \cite{ordentlich2014approximate}. We extend the nested lattice framework of \cite{ordentlich2014approximate} to ensure security in our scheme.\\
\indent The rest of the paper is organized as follows: in Sec. II, we formally state the problem, in Sec. III our main results are presented. Sec. IV is devoted to the achievability scheme along with proofs of reliability and security analysis. Finally, The paper is concluded in Sec. V.
\section{Problem Statement}\label{sec2}
We study the $K$-receiver Gaussian broadcast channel with confidential messages. The transmitter has $K$ confidential messages, $W_1, W_2, \dots, W_K$, for receivers $1, 2, \dots, K$, respectively. Each receiver acts as a passive eavesdropper with respect to all messages excluding its own intended message. In addition, there are $K$ helpers sending jamming signals to protect the confidentiality of messages at unintended receivers. The goal is to ensure the reliability of the intended messages and the confidentiality of the unintended messages. The relation among the transmitter's and the helpers' inputs and the output of the channel at receiver $\ell$ is determined as:
\begin{IEEEeqnarray}{l}\label{eqchannel}
\mathbf{y}_{\ell}=h_{\ell}\mathbf{x}+\sum_{i=1}^K g_{i\ell}\mathbf{x}^J_{i}+\mathbf{z}_{\ell}
\end{IEEEeqnarray} 
$\mathbf{y}_{\ell}$ is receiver $\ell$'s observation from the channel, $\mathbf{x}$ is the $N$-length input vector transmitted by the transmitter, $h_{\ell}$ is the main channel gain from the transmitter to receiver $\ell$. Moreover, $\mathbf{x}^J_{i}$ is the jamming signal transmitted by the $i$-th helper, $g_{i\ell}$ is the gain of the channel between helper $i$ and receiver $\ell$. Note that we consider real-valued channel gains in our model. Finally, $\mathbf{z}_{\ell}$ is an independent i.i.d. Gaussian noise with zero means and unit variances. The power constraints at the transmitter and the helpers are given as $
\|\mathbf{x}\|^2 \leq NP~\mathrm{and}~\sum_{i=1}^K\|\mathbf{x}^J_i\|^2 \leq NP$. The confidential message, $W_{\ell}$, is independent of all other messages and is uniformly distributed over the set $\{1,\dots, 2^{NR_{\ell}}\}$, for $\ell \in~\{1,\dots,K\}$. The transmitter maps the messages to codeword $\mathbf{x}$ through a stochastic encoder, i.e., $\mathbf{x} = \mathcal{E}(W_1,W_2,\dots,W_K)$. At receiver $\ell$, decoder $\mathcal{D}_{\ell}$ estimates the respective transmitted message as $\hat{W}_{\ell}=\mathcal{D}_{\ell}(\mathbf{y}_{\ell})$. Figure 1 illustrates the communication model.
\vspace{1mm}
\begin{defi}[Achievable secure rates]\label{achievability tuple}
For the $K$-receiver Gaussian broadcast channel with $K$ independent confidential messages, a non-negative rate tuple $(R_1,R_2,\dots, R_K)$ is achievable, if for any$~\epsilon>0$ and sufficiently large $N$, there exist encoder $\mathcal{E}$ and decoders $\{\mathcal{D}_{\ell}\}_{\ell=1}^{K}$ such that $\forall \ell \in \{1,\dots,K\}$:

\vspace{-1mm}
\small
\begin{equation}
\mathrm{Prob}\left(D_{\ell}(\mathbf{y}_{\ell})\neq W_{\ell}\right)< \epsilon \label{eqic3}
\end{equation}
\begin{equation}\label{eqic4}
R_{\ell}\leq \frac{1}{N}H(W_{\ell}|\mathbf{y}_1,\dots,\mathbf{y}_{\ell-1},\mathbf{y}_{\ell+1},\dots,\mathbf{y}_K)+\epsilon~~ 
\end{equation}
\normalsize
\end{defi}
Inequalities (\ref{eqic3}) and (\ref{eqic4}) capture the reliability and the confidentiality constraints of message $W_{\ell}$, respectively; the confidentiality constraint ensures weak secrecy~\cite{maurer2000information}. The secrecy capacity region is the supremum over all the achievable secure rate tuples.
\begin{figure}\label{fig1}
\centering
\begin{tikzpicture}[scale=0.73]
\draw(0.7,6) node (nodew1) {$W_1^K$};
\draw (2,6) node[draw] (nodeE1) {$\mathrm {Tx}$};
\draw (2,4) node[draw] (nodeEJ1) {$\mathcal E^J_1$};
\draw (2,3) node {$\vdots$};
\draw (2,1.5) node[draw] (nodeEJK) {$\mathcal E^J_K$};
\draw (3,4) node[circle,fill,inner sep=1] (nodexj1){};
\draw (3,6) node[circle,fill,inner sep=1] (nodex){};
\draw (3,1.5) node[circle,fill,inner sep=1] (nodexjk){};
\draw (7,6) node[circle,draw,inner sep=0] (nodeplus1) {$+$};
\draw (7,7) node (nodenoise1) {$\mathbf z_1$};
\draw (9,6) node[draw] (nodeD1) {$\mathcal D_1$};
\draw (7,5) node (nodedots4){$\vdots$};
\draw (7,4) node (nodedots5){$\vdots$};
\draw (7,3) node (nodedots6){$\vdots$};
\draw (9,5) node (nodedots1){$\vdots$};
\draw (9,4) node (nodedots2){$\vdots$};
\draw (9,3) node (nodedots3){$\vdots$};
\draw(10.7,6) node (nodewhat1) {$\begin{array}{c}
 \hat W_1 \end{array} $
 };
 \draw(11.5,6) node (nodewhat4) {$\begin{array}{c}
 W_2^K \end{array} $
 }; 
\draw (11.5,6) node[cross=8pt,red] {};
\draw[->] (nodew1)--(nodeE1);
\draw[->] (nodeE1)--(nodex)node[pos=0.5,sloped,above]{$\mathbf{x}$};
\draw[->] (nodeEJ1)--(nodexj1)node[pos=0.7,sloped,above]{$\mathbf{x^J_1}$};
\draw[->] (nodeEJK)--(nodexjk)node[pos=0.7,sloped,above]{$\mathbf{x^J_K}$};
\draw[->] (nodex)--(nodeplus1)  node[pos=0.68,sloped,above] {$\scriptstyle \mathrm h_{1}$};
\draw[->] (nodexj1)--(nodeplus1)  node[pos=0.68,sloped,above] {$\scriptstyle \mathrm g_{11}$};
\draw[->] (nodexjk)--(nodeplus1)  node[pos=0.68,sloped,above] {$\scriptstyle \mathrm g_{K1}$};
\draw[->] (nodenoise1)--(nodeplus1);
\draw[->] (nodeplus1)--(nodeD1) node[pos=0.3,sloped,above]{$\mathbf y_1$};
\draw[->] (nodeD1)--(nodewhat1);
\draw (7,1.5) node[circle,draw,inner sep=0] (nodeplus3) {$+$};
\draw[->] (nodex)--(nodeplus3) node[pos=0.2,sloped,above]{$\scriptstyle \mathrm h_{K}$};
\draw[->] (nodexj1)--(nodeplus3) node[pos=0.2,sloped,above]{$\scriptstyle \mathrm g_{1K}$};
\draw[->] (nodexjk)--(nodeplus3) node[pos=0.2,sloped,above]{$\scriptstyle \mathrm g_{KK}$};
\draw (7,2.4) node (nodenoise3) {$ \mathbf{z}_K$};
\draw[->] (nodenoise3) -- (nodeplus3){};
\draw (9,1.5) node[draw] (nodeD3) {$ \mathcal D_K $};
\draw[->] (nodeplus3)--(nodeD3) node[pos=0.3,sloped,above]{$\mathbf y_K$};
\draw(10.7,1.5) node (nodewhat2) {$\begin{array}{c}
 \hat W_K \end{array} $
 };
\draw(11.8,1.5) node (nodewhat3) {$\begin{array}{c}
 W_1^{K-1} \end{array} $
 }; 
\draw (11.5,1.5) node[cross=8pt,red] {};
\end{tikzpicture}
\vspace{0.02 in}
\caption{ \small {The $K$-receiver Gaussian broadcast channel with confidential messages and $K$ helpers.}}
\vspace{-5mm}
\end{figure}
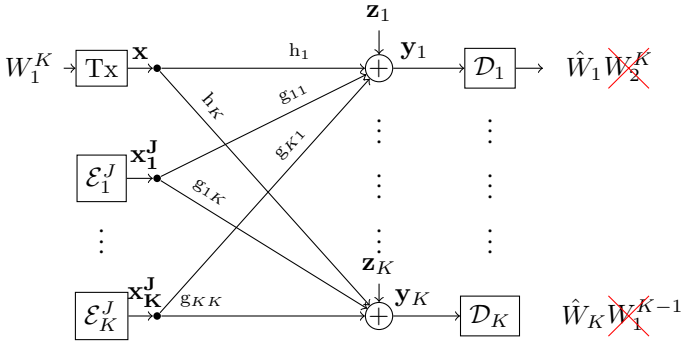
\section{Main results}\label{sec3}
We present our main result as a set of lower bounds on the individual secrecy rates of the confidential messages. We also present the sum secure degrees of freedom attainable by our scheme. 
\begin{theorem}
For the $K$-user Gaussian broadcast channel with $K$ helpers, any non-negative secure rate tuple $(R_1,R_2, \dots,R_K)$ satisfying the following inequalities is achievable with weak secrecy.
\vspace{-1mm}
\small
\begin{IEEEeqnarray}{l}\label{th1}
\medmuskip=-1.5mu
\thinmuskip=-1.5mu
\thickmuskip=-1.5mu
R_{\ell} < R_{comb,K}^{(\ell)}-
\frac{1}{2}\max_{k \in \{1,\dots,K\} \atop k \neq \ell}\log\left(\frac{\sum_{m=1}^M (h_k^2 P_{m \ell}+g_{\ell k}^2P_{m \ell}^ J)}{g_{\ell k}^2 P_{m^{\prime} \ell}^ J}\right)
\end{IEEEeqnarray} 
\normalsize
The rate $R_{comb,K}^{(\ell)}$ is defined as the optimal achievable rate at which receiver $\ell$ decodes the $K$-th linear integer combination using the compute-and-forward strategy. Also, $M$ is the number of dimensions used in the beam-forming operation and $m$ is the dimension index. $P_{m \ell}$ is the power allocated to encode the $m$-th component of the $\ell$-th confidential message. $P_{m \ell} ^J$ is the power used by helper $\ell$ to encode the $m$-th component of its jamming signal. Lastly, $P_{m^{\prime} \ell}^ J$ is the smallest power among the powers used to encode the components of the jamming signal by helper $\ell$, in our design this is also smaller than $P_{m \ell}$, $\forall m$. Also, the power allocated to encode helper $\ell$'s jamming signal is chosen such that $g_{\ell \ell}^2\sum_m P_{m\ell}^J <1$.\\  
\end{theorem}
\begin{remark}
The set of achievable rates in (\ref{th1}) for all $\ell \in \{1,\dots,K\}$ can be optimized over the choice of power allocations in the transmitter and the helpers as long as the following conditions are satisfied:
\vspace{-2mm}
\small
\begin{IEEEeqnarray}{l}
\medmuskip=-0.5mu
\thinmuskip=-0.5mu
\thickmuskip=0.51mu
 g_{\ell \ell}^2\sum_{m=1}^M P_{m\ell}^J <1,~~ P_{m_1\ell}^J <P_{m_2\ell} ~\forall m_1,m_2 \in [1,M],~ \forall \ell \in [1,K]\label{powerrestrict1}\\
 \sum_{\ell=1}^K\sum_{m=1}^M P_{m\ell}\leq P~, \sum_{\ell=1}^K\sum_{m=1}^M P_{m\ell}^J\leq P.\label{powerrestrict3}
\end{IEEEeqnarray}
\normalsize
\end{remark}
Our achievable scheme is based on rate-splitting, nested lattice coding, i.i.d. repetitions, cooperative jamming, and beam-forming. Figure 2 illustrates the block diagram of the encoding steps at the transmitter. The decoding is performed according to the asymmetric compute-and-forward strategy at the receivers. The detailed description is provided in Sec. \ref{sec4}.
\begin{figure}\label{fig2}
\centering
\includegraphics[width=0.4\textwidth]{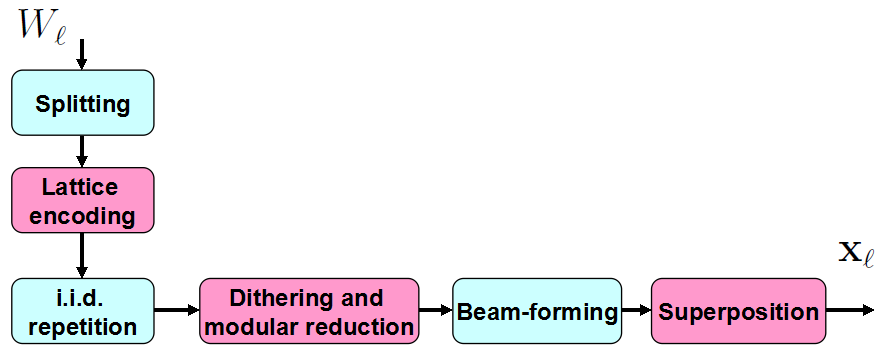}
\caption{\small{The encoding steps performed by the transmitter for each confidential message.}}
\end{figure}
\begin{corollary}
Following our scheme, at each receiver, a $\frac{1}{K}$ secure degrees of freedom is achievable for the receiver's intended message; hence, the optimal sum secure degrees of freedom of $1$ is achievable, i.e.,
\begin{equation}\label{cor1}
s.d.o.f.\triangleq \lim_{P \rightarrow \infty} \frac{\sum_{\ell=1}^KR_{\ell}}{\frac{1}{2}\log(1+P)}= 1
\end{equation} 
Corollary 1 is proven in Sec. \ref{sec4}.
\end{corollary}
\section{Achievability scheme}\label{sec4}
We describe our achievable scheme for $K=3$ receivers and three helpers to better clarify the key ideas in our coding scheme. Then, we generalize our scheme to any arbitrary $K>2$ receivers with $K$ helpers. We begin with codebook construction at the transmitter and then we describe the codebook construction at the helpers.
\subsection{Codebook construction}
\indent The transmitter generates a lattice vector for each independent confidential message. The lattice vectors are drawn from a set of nested lattice sets. \\
\indent Consider pairs of coarse and fine lattices as $(\Lambda^m_\ell,\Lambda^m_{f,\ell})$ for each pair $(m,\ell) \in \{1,\dots,T^4\}\times\{1,2,3\}$. Similarly, consider pairs $(\Lambda^m_{Ji},\Lambda^m_{f,Ji})$ for $i\in\{1,2,3\}$, and $m \in \{1,\dots,T^4\}$. The subscript $f$ specifies the fine lattice in the pair. $T$ is some large number; let us define $M\triangleq T^4$. Assume that these lattices are nested according to the following chain:
\begin{equation*}
\medmuskip=-1mu
\thinmuskip=-1mu
\thickmuskip=-1mu
\Lambda\subseteq \Lambda^m_3 \subseteq \Lambda^m_2 \subseteq \Lambda^m_1 \subseteq \Lambda^m_{3J} \subseteq \Lambda^m_{2J} \subseteq \Lambda^m_{1J} \subseteq \Lambda^m_{f,3} ...
\end{equation*}
\begin{equation}\label{chainlatt}
\medmuskip=-1mu
\thinmuskip=-1mu
\thickmuskip=-1mu
\subseteq \Lambda^m_{f,2} \subseteq \Lambda^m_{f,1} \subseteq \Lambda^m_{f,3J} \subseteq \Lambda^m_{f,2J} \subseteq \Lambda^m_{f,1J}
\end{equation}
The above coarse lattice sets are scaled such that their second moments are equal to $\sigma^2_{m3},\sigma^2_{m2},\sigma^2_{m1},\sigma^2_{m,3J},\sigma^2_{m,2J},\sigma^2_{m,1J}$, respectively. We denote the fundamental Voronoi region of the coarse lattice $\Lambda^m_\ell$ as $\mathcal{V}^m_{\ell}$; similarly, the fundamental Voronoi region of the coarse lattice $\Lambda^m_{iJ}$ is denoted as $\mathcal{V}^m_{iJ}$. The centers of the cosets of the fine lattice sets $\Lambda^m_{f,\ell}$ and $\Lambda^m_{f,iJ}$ are both n-length lattice words, which are the realizations of the n-length random vector $\mathbf{t}_{m\ell}$ and $\mathbf{u}_{mi}$, respectively. The inner codebook associated with sub-message $(m,\ell)$ is defined as $\mathcal{L}_{m\ell}\triangleq \{\mathbf{t}_{m\ell}|\mathbf{t}_{m\ell} \in \mathcal{V}^m_{\ell}\}$. Also, the inner codebook $\mathcal{L}_{m,Ji}$ is similarly defined for the collection of the jamming codewords $\mathbf{u}_{mi}$ and is used by the $i$-th helper.\\
\indent Consider a probability distribution $P(\mathbf{t}_{m\ell})$ over the codebook $\mathcal{L}_{m\ell}$. To generate the outer codebooks for sub-message $(m,\ell)$, the transmitter acts as follows: from codebook $\mathcal{L}_{m\ell}$ and according to distribution $P(\mathbf{t}_{m\ell})$, it draws $B$ i.i.d. copies of codewords $\mathbf{t}_{m\ell}$ and then, it concatenates the drawn vectors. The resulting codeword which has length $N\triangleq n \times B$ is considered as one realization of the outer codeword $\bar{\mathbf{t}}_{m\ell}$. The transmitter generates $2^{NR^{\ell}_{comb3,m}}$ realizations of random vector $\bar{\mathbf{t}}_{m\ell}$, where $R^{\ell}_{comb3,m}>0$ and $R^{\ell}_{comb,3}\triangleq\sum_{m=1}^MR^{\ell}_{comb3,m}$. The collection of the generated codewords is termed the outer codebook for sub-message $(m,\ell)$ and denoted as $\mathcal{C}_{m\ell}$. The outer codebook at helper $i$, generated in a similar manner, and denoted as $\mathcal{C}_{m,iJ}$. Note that the idea behind the i.i.d. repetitions of the inner codewords is to take advantage of the Packing lemma in the proof of weak secrecy. \footnote{Packing Lemma is deduced by applying the joint typicality lemma on i.i.d. random sequences \cite{el2011network}.}\\
\indent Next step in the codebook construction is the random partitioning. For each sub-message codebook $\mathcal{C}_{m\ell}$, the transmitter randomly partitions the outer codewords into $2^{NR_{m\ell}}$ bins of equal sizes. The transmitter chooses the non-negative rates $R_{m\ell}$ such that $R_\ell=\sum_{m=1}^M R_{m\ell}$, where 
\begin{IEEEeqnarray}{l}\label{rl3}
\medmuskip=0mu
\thinmuskip=0mu
\thickmuskip=0mu
\nonumber R_{\ell}\triangleq R_{comb,3}^{(\ell)}-\max_{k\in \{1,2,3\} \atop k \neq \ell}\left(\frac{1}{2}\log\left(\frac{\sum_m (h_k^2P_{m\ell}+g_{\ell k}^2P_{m\ell}^J)}{g_{\ell k}^2P_{m^{\prime}\ell}^J}\right)\right)\\
+\epsilon_{\ell},
\end{IEEEeqnarray}
in which term $\epsilon_{\ell}$ vanishes as the block length increases. \\
To each partition, an index $w_{m\ell} \in \{1, \dots, 2^{NR_{m\ell}}\}$ is randomly assigned. Additionally, for each sub-message $w_{m\ell}$, the transmitter generates a random dither vector $\mathbf{d}_{m\ell}$ drawn from a uniform distribution over the Voronoi region $\mathcal{V}^m_{\ell}$. The outer dither codewords $\bar{\mathbf{d}}_{m\ell}$ are constructed as described before. 
\subsection{Encoding} The transmitter encodes the confidential message $w_{\ell}$ by dividing the message into $M=T^4$ independent sub-messages, where $T$ is a large number. It is worth to mention that $M$ is the number of dimensions used in beam-forming the signals. Our ultimate goal is to align codewords at the unintended receivers with the jamming signals in many dimensions. Each sub-message is denoted by indices $(m,\ell)$, where $m \in \{1,\dots,M\}$ and $\ell \in \{1,\dots, K\}$, and is encoded separately. To encode the sub-message $w_{m\ell}$, the transmitter picks randomly a codeword $\bar{\mathbf{t}}_{m\ell}$ from the corresponding codebook $\mathcal{C}_{m\ell}$. It then dithers the extracted codeword and reduces the sum through a modular operation over the corresponding coarse lattice, i.e.,
\begin{equation}\label{firstmod}
\tilde{\mathbf{x}}_{m\ell}\triangleq\left[\bar{\mathbf{t}}_{m\ell}+\bar{\mathbf{d}}_{m\ell}\right]\modnml
\end{equation}
The modular operation in (\ref{firstmod}) is done block-wise for each block of length $n$. Next, we apply beam-forming such that each codeword $\tilde{\mathbf{x}}_{m\ell}$ scaled as $
\mathbf{x}_{m\ell}\triangleq \tilde{\mathbf{x}}_{m\ell}. f(m,\ell, \mathbf{h},\mathbf{g}_1,\mathbf{g}_2, \mathbf{g}_3)$, where $ \mathbf{h}\triangleq [h_1,h_2,h_3]^T$, $\mathbf{g}_i\triangleq [g_{i1},g_{i2},g_{i3}]^T$, and $f(.)$ is a mapping which takes the indices and the channel gains as inputs and outputs an scalar value. The mapping $f$, is chosen such that the codewords $ \mathbf{x}_{m\ell}$ for all $(m,\ell)$ are rationally independent for all channel gain vectors, except for a small Lebesgue measure. We will expand on the mapping $f$ shortly. The transmitter sends codeword $\mathbf{x}\triangleq \sum_{\ell=1}^K\mathbf{x}_{\ell}$, where $\mathbf{x}_{\ell}\triangleq\sum_{m=1}^M \mathbf{x}_{m\ell}$, across the channel. The power allocated to sub-codeword $(m,\ell)$ is defined as $
P_{m\ell}\triangleq \sigma^2_{m\ell}.|f(m,\ell, \mathbf{h},\mathbf{g}_1,\mathbf{g}_2, \mathbf{g}_3)|^2$. It is worth mentioning that the lattice sets in (\ref{chainlatt}) are scaled such that the power allocations $P_{m\ell}$ and $P_{m\ell}^J$ satisfy the constraints in (\ref{powerrestrict1}-\ref{powerrestrict3}). \\
\indent Encoding at helper $\ell$ is performed as follows: for each $m\in \{1,\dots,M\}$, it randomly picks a jamming codeword, $\bar{\mathbf{u}}_{m\ell}$, from codebook $C_{m,\ell J}$. It then dithers the codeword and performs the modular operation using lattice $\Lambda^m_{\ell J}$ to generate $\tilde{\mathbf{x}}^J_{m\ell}$. Next, it constructs codeword $\mathbf{x}^J_{m\ell}$ as $
\mathbf{x}^J_{m\ell}\triangleq \tilde{\mathbf{x}}^J_{m\ell}. f(m,\ell, \mathbf{h},\mathbf{g}_1,\mathbf{g}_2, \mathbf{g}_3)$. Eventually, helper $\ell$ transmits signal $\mathbf{x}^J_{\ell}=\sum_m\mathbf{x}^J_{m\ell}$ through the channel. We denote the power of the transmitted codeword by helper $\ell$ as $P_{\ell}^J$ which is defined as $P_{\ell}^J\triangleq\sum_{m=1}^M P^J_{m\ell}$ over index $m$, where $P^J_{m\ell}$ is defined similar to $P_{m\ell}$. The construction of the beam-forming function $f$ is performed such that the desired alignments are formed. For $K=3$ receivers, and for a given $\ell$, codeword $\mathbf{x}_{\ell}$ should get aligned with jamming codeword $\mathbf{x}^J_{\ell}$ at receivers $k \neq \ell$. For instance, codeword $\mathbf{x}_{1}$ needs to be aligned with jamming codeword $\mathbf{x}^J_{1}$ at receivers 2 and 3. This requires that the same pairs of codewords get aligned at multiple receivers, simultaneously. To this end, we take advantage of the asymptotic alignment technique, (introduced in \cite{MotahariRef16} and used in \cite{xieJournal2} for real-alignment), to align the $N$-dimensional lattice codewords. To do so, consider a one-to-one mapping $\phi^3:\{1,\dots,M\}\rightarrow\{1,\dots,T\}\times \{1,\dots,T\}\times \{1,\dots,T\}\times \{1,\dots,T\}$. We design the beam-forming function, $f$, for the three-receiver Gaussian broadcast channel with channel gain vectors $\mathbf{h},\mathbf{g}_1,\mathbf{g}_2, \mathbf{g}_3$ as 

\vspace{-2mm}
\small
\begin{IEEEeqnarray}{l}\label{beamforming31}
 f(m,1, \mathbf{h},\mathbf{g}_1,\mathbf{g}_2, \mathbf{g}_3)=h_2^{r_1}h_3^{r_2}g^{r_3}_{12}g^{r_4}_{13}\\\label{32}
f(m,2, \mathbf{h},\mathbf{g}_1,\mathbf{g}_2, \mathbf{g}_3)=h_1^{r_1}h_3^{r_2}g^{r_3}_{21}g^{r_4}_{23}\\
\label{33}
f(m,3, \mathbf{h},\mathbf{g}_1,\mathbf{g}_2, \mathbf{g}_3)=h_1^{r_1}h_2^{r_2}g^{r_3}_{31}g^{r_4}_{32}\\
(r_1,r_2,r_3,r_4)=\phi^3(m),~m \in \{1,\dots, M\}.
\end{IEEEeqnarray}
\normalsize
Following the results in \cite{MotahariRef16} and \cite{xieJournal2}, it can be shown that for large enough values of $M$ our beam-forming function asymptotically provides the desired alignments at the receivers simultaneously. In other words, the desired alignments between the pair of codewords happen in many dimensions which asymptotically yields perfect alignment.
\subsection{Decoding}
We describe decoding at receiver 1; other receivers act in a similar manner. Receiver 1 observes the following sequence from the channel:

\vspace{-3mm}
\small
\begin{IEEEeqnarray}{l}\label{rec1obs}
\tiny{
\nonumber \mathbf{y}_1=h_1\sum_{m=1}^M \mathbf{x}_{m1}+\sum_{m=1}^M (h_1\mathbf{x}_{m2}+g_{21}\mathbf{x}^J_{m2})}\\
\tiny {+\sum_{m=1}^M (h_1\mathbf{x}_{m3}+g_{31}\mathbf{x}^J_{m3})+ \sum_{m=1}^M g_{11}\mathbf{x}^J_{m1}+\mathbf{z}_1}
\end{IEEEeqnarray}
\normalsize
Due to beam-forming in (\ref{beamforming31})-(\ref{33}), the sub-message codewords associated with confidential message $W_2$ and the jamming sub-codewords of helper $2$ are aligned in the second term of (\ref{rec1obs}). Similarly, the corresponding sub-codewords in the third term of the equation (\ref{rec1obs}) are aligned. Moreover, according to (\ref{powerrestrict1}), power of the fourth term in (\ref{rec1obs}) falls below noise level; as it carries no useful information regarding the messages, receiver 1 treats the fourth term as noise. Consequently, receiver 1 decodes the following effective $3$-user multiple-access channel in which the power of the effective noise $\tilde{\mathbf{z}}_1$ is normalized. We have

\vspace{-4mm}
\small
\begin{IEEEeqnarray}{l}\label{3userMAcEff}
\medmuskip=0mu
\thinmuskip=0mu
\thickmuskip=0mu
\nonumber \tilde{\mathbf{y}}_1=\frac{h_1}{\sqrt{g^2_{11}P^J_1+1}}\sum_{m=1}^M \mathbf{x}_{m1}+\frac{1}{\sqrt{g^2_{11}P^J_1+1}}\sum_{m=1}^M (h_1\mathbf{x}_{m2}+g_{21}\mathbf{x}^J_{m2})\\
+\frac{1}{\sqrt{g^2_{11}P^J_1+1}}\sum_{m=1}^M (h_1\mathbf{x}_{m3}+g_{31}\mathbf{x}^J_{m3})+\tilde{\mathbf{z}}_1
\end{IEEEeqnarray}
\normalsize
Receiver 1 decodes three effective lattice codewords, i.e., $\mathbf{x}_{eff,1}\triangleq\sum_{m=1}^M \mathbf{x}_{m1}$, $\mathbf{x}_{eff,2}\triangleq\sum_{m=1}^M (h_1\mathbf{x}_{m2}+g_{21}\mathbf{x}^J_{m2})$, and $\mathbf{x}_{eff,3}\triangleq\sum_{m=1}^M (h_1\mathbf{x}_{m3}+g_{31}\mathbf{x}^J_{m3})$. In other words, the effective channel vector $\mathbf{h}_{eff,1}$ at receiver 1 is defined as 
$
\medmuskip=0mu
\thinmuskip=0mu
\thickmuskip=0mu
\mathbf{h}_{eff,1}\triangleq\left[\frac{h_1}{\sqrt{g^2_{11}P^J_1+1}},\frac{1}{\sqrt{g^2_{11}P^J_1+1}},\frac{1}{\sqrt{g^2_{11}P^J_1+1}}\right]^T$, and subsequently, the power scaling factor which determines the ratios of the power of effective codewords with respect to the power constraint is $
\medmuskip=0mu
\thinmuskip=0mu
\thickmuskip=0mu
\mathbf{b}_{eff,1}\triangleq \left[\sqrt{\frac{P_1}{P}},\sqrt{\frac{h_1^2P_2+g^2_{21}P^J_2}{P}},\sqrt{\frac{h_1^2P_3+g^2_{31}P^J_3}{P}}\right]^T$. According to the asymmetric compute-and-forward technique and Theorem 7 in \cite{ordentlich2014approximate}, receiver 1 finds the optimal set of linearly independent integer-valued coefficient vectors, which maximizes the achievable sum rate, to construct the integer combinations and then it decodes the integer combinations successively. We denote these vectors as $\mathbf{a}_1$, $\mathbf{a}_2$, and $\mathbf{a}_3$. Upon decoding the  first integer combination, the codeword belonging to the densest lattice inner codebook is decoded. Let us denote the first integer combination by vector $\mathbf{v}_1\triangleq \sum_{\ell=1}^3\mathbf{a}_1(\ell)\mathbf{x}_{eff,\ell}$. Receiver 1 decodes $\mathbf{v}_1$ as follows:

\vspace{-3mm}
\small
\begin{equation}
\left[\mathbf{s}_1\triangleq \beta_1 \tilde{\mathbf{y}}_1-\sum_{\ell=1}^3 \mathbf{a}_1(\ell)\bar{\mathbf{d}}_{\ell}\right] \modn
=\left[\mathbf{v}_1+\mathbf{z}_{eff,1}\right] \modn,
\end{equation}
\normalsize
in which the effective noise of the first integer combination is defined as $
\mathbf{z}_{eff,1}\triangleq \sum_{\ell=1}^3 \left(\beta_1\mathbf{h}_{eff,1}(\ell)-\mathbf{a}_1(\ell)\right)\mathbf{x}_{eff,\ell}+\beta_1\tilde{\mathbf{z}}_1$. To decode the integer combination $\mathbf{v}_1$, receiver 1 computes the quantization value of $\mathbf{s}_1$ under the densest lattice among the lattice sets used for encoding $\{\mathbf{x}_{eff,\ell}\}_{\ell=1}^3$. Let us denote the index of the corresponding effective codeword with $k$. Then, according to Theorem 2 in \cite{ordentlich2014approximate} we have: $R^{1}_{comb,1}\triangleq\frac{1}{2}\log(\frac{P_{eff,k}}{\sigma^2_{eff,1}})$. $R^{1}_{comb,1}$ is the optimal achievable rate at which the first integer combination is decoded at receiver 1. Similarly, we can define $R^{1}_{comb,2}$ and $R^{1}_{comb,3}$ as the optimal rates of decoding the second and the third integer combinations at receiver 1, respectively. $P_{eff,k}$ is the power of the $k$-th effective codeword and $\sigma^2_{eff,1}$ is the variance of the effective noise associated with the first integer combination, i.e., $\mathbf{z}_{eff,1}$. Receiver 1 proceeds with decoding the next integer combinations of the effective codewords. However, to maximize the achievable rates, receiver 1 first cancels out the contribution of the previously decoded codewords from the current combination and then the codeword with the highest rate among the remaining codewords in the integer combination gets decoded. Assume that the effective codewords are decoded in the order specified by $\pi^{-1}(1),\pi^{-1}(2), \pi^{-1}(3)$, where $\pi(.)$ is a one-to-one permutation operator over the set $\{1,2,3\}$. Then, following Theorem 2 in \cite{ordentlich2014approximate}, the $k$-th optimal achievable combination rate is given as $
R^{1}_{comb,k}\triangleq\frac{1}{2}\log\left(\frac{P_{eff,\pi^{-1}(k)}}{\sigma^2_{eff,k}}\right)$, where $\sigma^2_{eff,k}$ is the variance of the effective noise in $k$-th integer combination. Note that upon decoding each combination, the effective codeword which was constructed using the densest lattice (highest rate) among the participating codewords in the combination is decoded. Therefore, the order among the variances of the effective noises is given as $\sigma^2_{eff,1}\leq \sigma^2_{eff,2}\leq \sigma^2_{eff,3}$. Note that the goal is to obtain a lower-bound on the achievable rate of $\mathbf{x}_{eff,1}$. According to the definition of $R^{1}_{comb,k}$ and the order among the variances, $R_{eff,1}\geq R^1_{comb,3} $. Therefore, receiver 1 can reliably decode its intended codeword so long as it is generated at a rate $R_{eff,1}\leq \frac{1}{2}\log(\frac{P_1}{\sigma^2_{eff,3}})$. Note that, given the optimal integer-valued coefficient vectors, i.e., $\mathbf{a}_1,\mathbf{a}_2,\mathbf{a}_3$, the variance $\sigma^2_{eff,3}$ is a function of $\beta_3$. It can be shown that the optimal choice for $\beta_3$ which minimizes $\sigma^2_{eff,3}$ is the MSE factor \cite{ordentlich2014approximate}, i.e, $
\beta_3=\frac{\mathbb{E}\left[\left(\sum_{\ell=1}^3\mathbf{a}_3(\ell)\mathbf{x}_{eff,\ell}\right)~\tilde{\mathbf{y}}_1\right]}{\mathbb{E}\left[\tilde{\mathbf{y}}^2_1\right]}
$. The integer-valued coefficients can be computed using the LLL reduction algorithm in \cite{LenstraLenstraLovasz1982} which provides a polynomial-time solution and computes a nearly optimal set of integer-valued coefficient vectors. \footnote{Due to space limitation, we will include numerical results in the extended version of this paper.} The proof of reliability at other receivers can be done similarly. So far, we showed that for a confidential message $\ell$, any non-negative rate below $ R^{\ell}_{comb,3}$ can be decoded reliably at receiver $\ell$ which ensures the reliability of the rates in Theorem 1. Next section is devoted to the analysis of security.
\subsection{Security analysis}\label{sec5}
In this section, we show that our achievable scheme provides weak secrecy for all messages at the unintended receivers, i.e.,
\vspace{-2mm}
\small
\begin{equation}\label{weaksec1}
\medmuskip=0mu
\thinmuskip=0mu
\thickmuskip=0mu
\frac{1}{nB}I(W_1,\dots, W_{\ell-1},W_{\ell+1},\dots,W_K;\mathbf{y}_{\ell})\leq \epsilon,~\forall\ell\in\{1,\dots,K\}
\end{equation}
\normalsize
in which $\epsilon>0$ tends to zero as $n$ and $B$ approach infinity. For simplicity, we shall prove (\ref{weaksec1}) for $K=3$ receivers; the extension of the proof to an arbitrary $K>2$ is straightforward. We proceed the proof by showing the weak secrecy of the joint messages $(W_2,W_3)$ at receiver 1, i.e., $\frac{1}{nB}I(W_2,W_3;\mathbf{y}_{1})\leq \epsilon$. We have $\frac{1}{nB}I(W_2,W_3;\mathbf{y}_{1})\leq \frac{1}{nB}I(W_2,W_3;\mathbf{y}_{1},\bar{\mathbf{t}}_1)$, therefore,

\vspace{-4mm}
\scriptsize
\begin{IEEEeqnarray}{l}\label{eqsec1}
\frac{1}{nB}I(W_2,W_3;\mathbf{y}_{1})\leq \sum_{\ell=2}^3 R_{\ell}- \frac{1}{nB}H(W_2,W_3|\mathbf{y}_{1},\bar{\mathbf{t}}_1),
\end{IEEEeqnarray}
\normalsize
in which $\bar{\mathbf{t}}_{\ell}\triangleq(\bar{\mathbf{t}}_{1\ell},\dots,\bar{\mathbf{t}}_{m\ell},\dots,\bar{\mathbf{t}}_{M\ell})$. We proceed by lower bounding the second term in (\ref{eqsec1}):

\scriptsize
\begin{IEEEeqnarray}{l}\label{eqsec2}
\small
\medmuskip=0mu
\thinmuskip=0mu
\thickmuskip=0mu
\nonumber \frac{1}{nB}H(W_2,W_3|\mathbf{y}_{1},\bar{\mathbf{t}}_1)=\frac{1}{nB}H(W_2,W_3,\bar{\mathbf{t}}_2,\bar{\mathbf{t}}_3|\mathbf{y}_{1},\bar{\mathbf{t}}_1)\\
\nonumber -\frac{1}{nB}H(\bar{\mathbf{t}}_2,\bar{\mathbf{t}}_3|\mathbf{y}_{1},\bar{\mathbf{t}}_1,W_2,W_3)\\
\nonumber
\geq \frac{1}{nB}H(\bar{\mathbf{t}}_2,\bar{\mathbf{t}}_3|\mathbf{y}_{1},\bar{\mathbf{t}}_1)
 -\frac{1}{nB}H(\bar{\mathbf{t}}_2,\bar{\mathbf{t}}_3|\mathbf{y}_{1},\bar{\mathbf{t}}_1,W_2,W_3)\\\nonumber
 \stackrel{(a)}{\geq} \frac{1}{nB}H(\bar{\mathbf{t}}_2,\bar{\mathbf{t}}_3|\mathbf{y}_{1},\bar{\mathbf{t}}_1) - 2\epsilon_{23}\\
 \nonumber
  \stackrel{(b)}{\geq} \frac{1}{nB}H(\bar{\mathbf{t}}_2,\bar{\mathbf{t}}_3|\mathbf{y}_{1},\bar{\mathbf{t}}_1,D,\mathbf{z}_1) - 2\epsilon_{23}\\
  \nonumber
  \medmuskip=-2mu
\thinmuskip=-2mu
\thickmuskip=-2mu
 \stackrel{(c)}{=}\frac{1}{nB}H\left(\bar{\mathbf{t}}_2,\bar{\mathbf{t}}_3\bigg|\sum_{m=1}^M (h_1\mathbf{x}_{m2}+g_{21}\mathbf{x}^J_{m2}),\sum_{m=1}^M (h_1\mathbf{x}_{m3}+g_{31}\mathbf{x}^J_{m3}),\bar{\mathbf{t}}_1,D,\mathbf{z}_1\right)
 - 2\epsilon_{23}\\   
 \nonumber
  \medmuskip=-2mu
\thinmuskip=-2mu
\thickmuskip=-2mu
 \stackrel{(d)}{=}\frac{1}{nB}H\left(\bar{\mathbf{t}}_2,\bar{\mathbf{t}}_3\bigg|\sum_{m=1}^M (h_1f_{m2}\bar{\mathbf{t}}_{m2}+g_{21}f_{m2}\bar{\mathbf{u}}_{m2}),\right.\\
 \nonumber
  \medmuskip=-1mu
\thinmuskip=-1mu
\thickmuskip=-1mu
 \left. \sum_{m=1}^M (h_1f_{m3}\bar{\mathbf{t}}_{m3}+g_{31}f_{m3}\bar{\mathbf{u}}_{m3}),\bar{\mathbf{t}}_1,D,\mathbf{z}_1\right)
 - 2\epsilon_{23}\\ 
 \medmuskip=-1mu
\thinmuskip=-1mu
\thickmuskip=-1mu
\stackrel{(e)}{=}\frac{1}{nB}H\left(\bar{\mathbf{t}}_2,\bar{\mathbf{t}}_3\bigg|\sum_{m=1}^M (\tilde{\mathbf{t}}_{m2}+\tilde{\mathbf{u}}_{m2}),\sum_{m=1}^M (\tilde{\mathbf{t}}_{m3}+\tilde{\mathbf{u}}_{m3}),\bar{\mathbf{t}}_1,D,\mathbf{z}_1\right)- 2\epsilon_{23}\label{eqsec3}
\end{IEEEeqnarray}
\normalsize
In the above arguments, inequality (a) holds due to Lemma 1 in \cite{babaheidarian2}. Inequality (b) is true since conditioning reduces entropy. Equality (c) is deduced from expression (\ref{rec1obs}) and definition of $\mathbf{x}_{m\ell}$. Equality (d) is deduced from (\ref{firstmod}) and after subtracting dithers. Also, equality (e) comes from defining the lattice vectors $h_1f_{m2}\bar{\mathbf{t}}_{m2}$, $g_{21}f_{m2}\bar{\mathbf{u}}_{m2}$, $h_1f_{m3}\bar{\mathbf{t}}_{m3}$, and $g_{31}f_{m3}\bar{\mathbf{u}}_{m3}$ as lattice vectors $\tilde{\mathbf{t}}_{m2}$, $\tilde{\mathbf{u}}_{m2}$, $\tilde{\mathbf{t}}_{m3}$, and $\tilde{\mathbf{u}}_{m3}$, respectively.  \\
\indent Now, assume that among the nested coarse lattices $\{\Lambda_{m2}\}^M_{m=1}$ and $\{\Lambda^J_{m2}\}^M_{m=1}$, lattice $\Lambda^J_{m^{\prime}2}$ is the densest lattice for some $m^{\prime} \in \{1,\dots,M\}$ and similarly, assume among the nested coarse lattices $\{\Lambda_{m2}\}^M_{m=1}$ and $\{\Lambda^J_{m3}\}^M_{m=1}$, lattice $\Lambda^J_{m^{\prime \prime}3}$ is the densest lattice for some $m^{\prime \prime}\in \{1,\dots,M\}$. Then, following the expression in (\ref{eqsec3}), we have: 
\vspace{-2mm}

\scriptsize
\begin{IEEEeqnarray}{l}
\medmuskip=-1mu
\thinmuskip=-1mu
\thickmuskip=-1mu
\nonumber \frac{1}{nB}H(W_2,W_3|\mathbf{y}_{1},\bar{\mathbf{t}}_1)
\geq \frac{1}{nB}H\left(\bar{\mathbf{t}}_2,\bar{\mathbf{t}}_3\bigg|\left[\sum_{m=1}^M (\tilde{\mathbf{t}}_{m2}+\tilde{\mathbf{u}}_{m2})\right]\modmprimes,\right.\\
\nonumber
\medmuskip=-1mu
\thinmuskip=-1mu
\thickmuskip=-1mu
\left. Q_{\Lambda^J_{m^{\prime}2}}(\sum_{m=1}^M (\tilde{\mathbf{t}}_{m2}+\tilde{\mathbf{u}}_{m2})), \left[\sum_{m=1}^M (\tilde{\mathbf{t}}_{m3}+\tilde{\mathbf{u}}_{m3})\right]\modmprimet, \right.\\
\nonumber
\medmuskip=-1mu
\thinmuskip=-1mu
\thickmuskip=-1mu
\left.Q_{\Lambda^J_{m^{\prime \prime}3}}(\sum_{m=1}^M (\tilde{\mathbf{t}}_{m3}+\tilde{\mathbf{u}}_{m3})),\bar{\mathbf{t}}_1,D,\mathbf{z}_1\right)- 2\epsilon_{23}\\
\nonumber
\medmuskip=-2mu
\thinmuskip=-2mu
\thickmuskip=-2mu
\geq \frac{1}{nB}H\left(\bar{\mathbf{t}}_2,\bar{\mathbf{t}}_3\bigg|\left[\sum_{m=1}^M (\tilde{\mathbf{t}}_{m2}+\tilde{\mathbf{u}}_{m2})\right]\modmprimes, \left[\sum_{m=1}^M (\tilde{\mathbf{t}}_{m3}+\tilde{\mathbf{u}}_{m3})\right]\modmprimet\right)\\
\nonumber
\medmuskip=-2mu
\thinmuskip=-2mu
\thickmuskip=-2mu
-\frac{1}{nB}H\left(Q_{\Lambda^J_{m^{\prime}2}}(\sum_{m=1}^M (\tilde{\mathbf{t}}_{m2}+\tilde{\mathbf{u}}_{m2})), \right. \\\nonumber
\left. Q_{\Lambda^J_{m^{\prime \prime}3}}(\sum_{m=1}^M (\tilde{\mathbf{t}}_{m3}+\tilde{\mathbf{u}}_{m3}))\bigg| \bar{\mathbf{t}}_1,D,\mathbf{z}_1\right)
- 2\epsilon_{23}\\
\nonumber
\medmuskip=-2mu
\thinmuskip=-2mu
\thickmuskip=-2mu
\stackrel{(f)}{\geq} \frac{1}{nB}H\left(\bar{\mathbf{t}}_2,\bar{\mathbf{t}}_3\bigg|\left[\sum_{m=1}^M (\tilde{\mathbf{t}}_{m2}+\tilde{\mathbf{u}}_{m2})\right]\modmprimes,\right.\\
\nonumber
\medmuskip=-2mu
\thinmuskip=-2mu
\thickmuskip=-2mu
\left. \left[\sum_{m=1}^M (\tilde{\mathbf{t}}_{m3}+\tilde{\mathbf{u}}_{m3})\right]\modmprimet\right)- 2\epsilon_{23}\\\nonumber
\medmuskip=-2mu
\thinmuskip=-2mu
\thickmuskip=-2mu
-\frac{1}{nB}H\left(Q_{\Lambda^J_{m^{\prime}2}}(\sum_{m=1}^M (\tilde{\mathbf{t}}_{m2}+\tilde{\mathbf{u}}_{m2})),Q_{\Lambda^J_{m^{\prime \prime}3}}(\sum_{m=1}^M (\tilde{\mathbf{t}}_{m3}+\tilde{\mathbf{u}}_{m3}))\right)\\
\nonumber
\medmuskip=-2mu
\thinmuskip=-2mu
\thickmuskip=-2mu
\geq\frac{1}{nB}H\left(\bar{\mathbf{t}}_2,\bar{\mathbf{t}}_3\bigg|\left[\sum_{m=1}^M (\tilde{\mathbf{t}}_{m2}+\tilde{\mathbf{u}}_{m2})\right]\modmprimes, \left[\sum_{m=1}^M (\tilde{\mathbf{t}}_{m3}+\tilde{\mathbf{u}}_{m3})\right]\modmprimet\right)\\
\nonumber
\medmuskip=-2mu
\thinmuskip=-2mu
\thickmuskip=-2mu
-\frac{1}{nB}H\left(Q_{\Lambda^J_{m^{\prime}2}}(\sum_{m=1}^M (\tilde{\mathbf{t}}_{m2}+\tilde{\mathbf{u}}_{m2}))\right)- 2\epsilon_{23}-\frac{1}{nB}H\left(Q_{\Lambda^J_{m^{\prime \prime}3}}(\sum_{m=1}^M (\tilde{\mathbf{t}}_{m3}+\tilde{\mathbf{u}}_{m3}))\right)\\
\nonumber
\medmuskip=-2mu
\thinmuskip=-2mu
\thickmuskip=-2mu
\stackrel{(g)}{\geq} \frac{1}{nB}H\left(\bar{\mathbf{t}}_2,\bar{\mathbf{t}}_3\bigg|\left[\sum_{m=1}^M (\tilde{\mathbf{t}}_{m2}+\tilde{\mathbf{u}}_{m2})\right]\modmprimes,\right. \\
\nonumber
\medmuskip=-2mu
\thinmuskip=-2mu
\thickmuskip=-2mu
\left. \left[\sum_{m=1}^M (\tilde{\mathbf{t}}_{m3}+\tilde{\mathbf{u}}_{m3})\right]\modmprimet\right)
-\frac{1}{2}\log\left(\frac{\sum_m (h_1^2P_{m2}+g_{21}^2P_{m2}^J)}{g_{21}^2P_{m^{\prime}2}^J}\right)\\\nonumber
\medmuskip=-2mu
\thinmuskip=-2mu
\thickmuskip=-2mu
-\frac{1}{2}\log\left(\frac{\sum_m (h_1^2P_{m3}+g_{31}^2P_{m3}^J)}{g_{31}^2P_{m^{\prime \prime} 3}^J}\right)-\delta(\epsilon_2)-\delta(\epsilon_3)-2\epsilon_{23}\\
\nonumber
\medmuskip=-2.2mu
\thinmuskip=-2.2mu
\thickmuskip=-2.2mu
\stackrel{(h)}{=}
\frac{1}{nB}H\left(\bar{\mathbf{t}}_2,\bar{\mathbf{t}}_3\right)
-\frac{1}{2}\log\left(\frac{\sum_m (h_1^2P_{m2}+g_{21}^2P_{m2}^J)}{g_{21}^2P_{m^{\prime}2}^J}\right)\\
\nonumber
\medmuskip=-2.2mu
\thinmuskip=-2.2mu
\thickmuskip=-2.2mu
-\frac{1}{2}\log\left(\frac{\sum_m (h_1^2P_{m3}+g_{31}^2P_{m3}^J)}{g_{31}^2P_{m^{\prime \prime} 3}^J}\right)-\delta(\epsilon_2)-\delta(\epsilon_3)- 2\epsilon_{23}\\
\nonumber
\medmuskip=-2.2mu
\thinmuskip=-2.2mu
\thickmuskip=-2.2mu
\stackrel{(k)}{=}
R_{comb,3}^{(2)}+R_{comb,3}^{(3)}-\frac{1}{2}\log\left(\frac{\sum_m (h_1^2P_{m2}+g_{21}^2P_{m2}^J)}{g_{21}^2P_{m^{\prime}2}^J}\right)\\
\medmuskip=-2.2mu
\thinmuskip=-2.2mu
\thickmuskip=-2.2mu
-\frac{1}{2}\log\left(\frac{\sum_m (h_1^2P_{m3}+g_{31}^2P_{m3}^J)}{g_{31}^2P_{m^{\prime \prime} 3}^J}\right)-\delta(\epsilon_2)-\delta(\epsilon_3)- 2\epsilon_{23}\label{eqsec6}
\end{IEEEeqnarray}
\normalsize
In the above inequalities, inequality (f) holds since conditioning reduces entropy. Inequality (g) is deduced by applying Lemma 1 in \cite{paper1} to lattice codewords $\sum_{m=1}^M (\tilde{\mathbf{t}}_{m2}+\tilde{\mathbf{u}}_{m2}))$ and $\sum_{m=1}^M (\tilde{\mathbf{t}}_{m3}+\tilde{\mathbf{u}}_{m3}))$. Equality (h) is deduced from Crypto Lemma in \cite{forney2003role} and the fact that the lattice sets used for encoding the jamming signals were chosen such that it would be denser than the lattice sets used for encoding the message signals. Finally, equality (k) is resulted from the independence of codewords $\bar{\mathbf{t}}_2$ and $\bar{\mathbf{t}}_3$ and the rates at which they were generated according to the achievable scheme in Section \ref{sec4}. Next, we plug the lower bound in (\ref{eqsec6}) to the second term in (\ref{eqsec1}). Then, following (\ref{rl3}) we obtain: $\frac{1}{nB}I(W_2,W_3;\mathbf{y}_{1})\leq \delta(\epsilon_2)+\delta(\epsilon_3)+2\epsilon_{23}+\epsilon_2+\epsilon_3$. Now, define $\epsilon^{\prime} \triangleq \delta(\epsilon_2)+\delta(\epsilon_3)+2\epsilon_{23}+\epsilon_2+\epsilon_3$, which tends to zero as $n$ and $B$ approach infinity. Thus, the analysis of weak secrecy for the joint messages $(W_2,W_3)$ at receiver 1 is completed. Proofs of weak secrecy for the unintended message pairs at receiver 2 and receiver 3 are established similarly.  \hfill $\blacksquare$
\vspace{-1mm}
\subsection*{Extension to an arbitrary $K>2$}
For the general case of $K>2$, the codebook construction is performed similar to $K=3$ case. However, in this case, each message is divided into $M\triangleq T^{2K-2}$ independent sub-messages where $T$ is some large number. Also, secure rates $R_{\ell}$ for $\ell \in \{1,\dots,K\}$ are chosen as $
R_{\ell}=R_{comb,K}^{(\ell)}-
\frac{1}{2}\max_{k \in \{1,\dots,K\} \atop k \neq \ell}\left(\log\left(\frac{\sum_{m=1}^M (h_k^2 P_{m \ell}+g_{\ell k}^2P_{m \ell}^ J)}{g_{\ell k}^2 P_{m^{\prime} \ell}^ J}\right)\right)+\epsilon_{\ell}$, where $\epsilon_{\ell}>0$ is a small number that vanishes as $N\rightarrow \infty$.\\
\indent Also, the encoding step is performed similar to $K=3$ case. The beam-forming functions used at the transmitter and at helpers are extended as in the following:

\vspace{-2mm}
\small
\begin{IEEEeqnarray}{l}\label{beamformingK}
\medmuskip=0mu
\thinmuskip=0mu
\thickmuskip=0mu
 \nonumber f(m,\ell, \mathbf{h},\mathbf{g}_1,\mathbf{g}_2, \dots, \mathbf{g}_K)= h_1^{r_1}h_2^{r_2}\dots h_{\ell-1}^{r_{\ell-1}}h_{\ell+1}^{r_{\ell}}\dots h_{K}^{r_{K-1}}\\
\medmuskip=0mu
\thinmuskip=0mu
\thickmuskip=0mu
\times g^{r_K}_{\ell 1}g^{r_{K+1}}_{\ell 2}\dots g^{r_{K+\ell-2}}_{\ell \ell-1} g^{r_{K+\ell-1}}_{\ell \ell+1} \dots g^{r_{2K-2}}_{\ell K},
\end{IEEEeqnarray}
\normalsize
where $(r_1,r_2,\dots,r_{2K-2})=\phi^K(m)$, in which $\phi^K(.)$ is a one-to-one mapping from the set $\{1,\dots,M\}$ to set of tuples with $2K-2$ elements, i.e., $(r_1,r_2,\dots,r_{2K-2})$, whose elements take values from the set $\{1,\dots,T\}$. Also, in (\ref{beamformingK}), vector $\mathbf{g}_{\ell}$ is defined as $\mathbf{g}_{\ell}\triangleq [ g_{\ell 1},g_{\ell 2}, \dots, g_{\ell K}]^T$. Decoding at each receiver is performed using the asymmetric compute-and-forward framework as in the case of $K=3$; the difference here is that each receiver decodes an effective $K$-user MAC to estimate its intended messages. Also, the the weak secrecy proof is a straightforward extension of $K=3$ case.
\subsection*{Proof of Corollary 1:}
The soundness of Corollary 1 is proven in two steps: step 1 is to show that the second term in (\ref{th1}) is constant with respect to power constraint $P$. Note that $P_{m\ell}$ and $P_{m\ell}^J$ are portions of powers allocated for transmitting the $\ell$-th confidential message by the transmitter and the jamming signal by helper $\ell$, respectively. Note that the power allocation must be performed in such a way that it satisfies power constraints in (\ref{powerrestrict1}-\ref{powerrestrict3}). Hence, we have $P_{m\ell}=\alpha_{m\ell} P$ and $P_{m\ell}=\alpha^J_{m\ell} P$, for some constants $0<\alpha_{m\ell}, \alpha^J_{m\ell} <1$. As a result, the second term in (\ref{th1}) can be rewritten as $
\frac{1}{2}\max_{k}\left(\log\left(\frac{P(\sum_m h_k^2\alpha_{m\ell}+g^2_{\ell k}\alpha^J_{m\ell})}{P g_{\ell k}^2 \alpha^J_{m^*\ell}}\right)\right)$. Notice that the factor $P$ would be canceled out from the top and bottom of the fraction and the rest is a constant with respect to power $P$. In step 2, we show that the first term in (\ref{th1}) provides $\frac{1}{K}$ degrees of freedom. Hence, total secure degrees of freedom provided by all confidential messages is $\sum_{\ell=1}^K\frac{1}{K}=1$. Note that in (\ref{th1}), $R_{comb,K}^{(\ell)}$ is the smallest combination rate among the optimal set of $K$ combination rates for the effective $K$-user MAC that receiver $\ell$ perceives. It was shown in Corollary 5 in \cite{ordentlich2014approximate} that for almost every channel gain vector, the degrees of freedom provided
by each of the $K$ optimal combination rates is $\frac{1}{K}$. Thus, $R_{comb,K}^{(\ell)}$ provides $\frac{1}{K}$ degrees of freedom as well and this holds for all $\ell \in \{1,\dots,K\}$. As a result, total secure degrees of freedom provided in our achievable scheme is equal to $1$, this is indeed optimal. Note that for a Gaussian broadcast channel with confidential messages $s.d.o.f. \leq 1$, since the optimal degrees of freedom for a Gaussian broadcast channel without security constraints is $1$ which serves as an upper bound in our security scenario \cite{xie2014secure}.
\vspace{-1mm}
\section{Conclusion}
We investigated transmitting confidential messages through the Gaussian broadcast channel with $K>2$ receivers and $K$ helpers. We offered an achievable scheme which achieves secure rates that operate within a constant gap from sum secure capacity.
\vspace{-3mm}
\bibliography{refn}
\end{document}